\documentclass{aa}  
\usepackage{graphicx}
\usepackage{txfonts}
\usepackage{natbib}
\begin{document}
   \title{On the nature of the cool component of MWC 560}


   \author{M. Gromadzki
          \inst{1}
          \and
          J. Miko{\l}ajewska\inst{1}
		  \and
		  P.A. Whitelock\inst{2,3}
		  \and
		  F. Marang\inst{2}
          }

  \offprints{M. Gromadzki}
 \mail{marg@camk.edu.pl}

   \institute{N. Copernicus Astronomical Center, Bartycka 18, PL-00-716 Warsaw, Poland     
         \and
             South African Astronomical Observatory, P.O. Box 9, Observatory, 7935, South 
			 Africa
       \and
National Astrophysics and Space Science Programme, Department of Mathematics
and Applied Mathematics, and the Department of Astronomy, University of Cape
Town, South Africa
             }

   \date{Received; accepted ??}

 
  \abstract
   {MWC 560 (V694~Mon) is one of the most enigmatic symbiotic system with a
   very active accretion-powered hot component. Such activity can be
   supported only by a luminous asymptotic giant branch star, i.e. a Mira or SR
   variable, with a high mass-loss rate.  It is also a very unusual jet
   source because the jet axis lies practically parallel to the line of
   sight. }
   {The aims of our study are the determination of the evolutionary status of the cool component of MWC 560.}
   {Our methods involve analysis of near-IR $JHKL$ and optical light curves.}
   {The cool component of MWC 560 pulsates with a period of $\sim$340~days,
   and it is probably a red SR variable on the thermally
   pulsing AGB. The high mass-loss rate expected for such a star is 
   sufficient to power the observed activity of the hot companion.}
   {}

   \keywords{stars: binaries: symbiotic  -- stars: long-period variables -- stars: mass loss -- stars: individual: MWC 560 = V694 Mon
-- infrared: stars}

   \maketitle
%

\section{Introduction}

Discovered by \citet{Merrill1943} as a star with strong emission lines,  MWC
560 (V694~Mon) is an enigmatic symbiotic system with a very active hot
component.  Its optical spectrum is always characterized by highly variable
absorption features, blue-shifted by 1000-6000 $\rm km\,s^{-1}$. These
originate from \ion{H}{i}, \ion{He}{i}, \ion{Ca}{ii} and \ion{Fe}{ii}, and
are detached from the stationary, narrow emission lines.  The blue-shifted
absorption features can be explained by a jet outflow along the line of
sight \citep{Toma1990}. In 1990, MWC 560 underwent a 2 mag photometric
outburst, accompanied by spectacular changes in these blue shifted
absorption lines \citep[e.g.][ and references therein]{Toma1990,
Schmid2001}. The hot component is also a permanent source of rapid flickering
with amplitudes in the range 0.2-0.7 mag on time scales of 10-100 min
\citep[e.g.][ and references therein]{grom2006}.

Both the flickering and the 1990 outburst characteristics suggest that the hot 
component is predominantly powered by unstable accretion, and its luminosity  
$\sim$100-1000 $L_{\odot}$ \citep{Toma1996, Mamiko1998} requires relatively
high accretion rates of  $\dot{M} \ga 10^{-7} M_{\odot} \mathrm{yr}^{-1}$
\citep{Schmid2001}.

\citet{Doroshenko1993} found a 1930-day period in the $m_{\rm{ph}}$ historical
light curve \citep{Lut1991} combined with more recent $B$-band photometry.
They suggested an orbital origin of this period: if the orbit is eccentric,
near periastron the accretion rate increases, and the hot component gets
brighter.

We know rather little about the cool component of MWC 560. However, the high
activity of its companion requires a high mass-loss rate, consistent with an
evolved AGB star rather than a normal red giant. Absorption features in the
optical spectrum suggest an M3$-$M4 spectral type \citep{Sand1973,
Allen1978}. Infrared spectral classification mainly based on TiO bands,
and the CaII triplet indicates an M4$-$M5 giant
\citep{Szkody1990,Bopp1990,ThiW1992}. Based on the lack of VO bands at 1.05
$\mu$m \citet{Meier1996} classified the cool component as an M5$-$M6 III giant. 
The most recent classification based on five TiO bands in the near infrared
resulted in M5.5 and M6 (two different observations) \citep{Murset1999}.
\citet{Franckowiak2003} found possible pulsations of red giant with $P=161$
days, and concluded that it is an AGB star.

This paper contains an analysis of near infrared ($JHKL$) and visual (AAVSO
and ASAS) light curves with the objective of searching for periodic changes
and establishing the nature of the cool giant.


\section{Observations}

%
   \begin{figure}
   \centering
   \includegraphics[width=8.5cm]{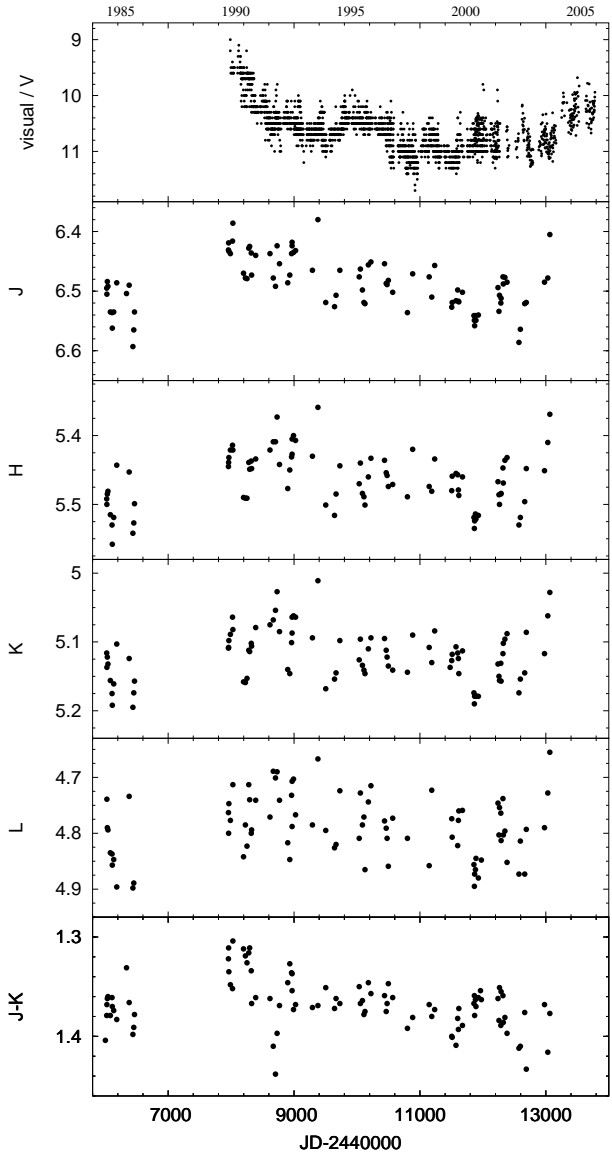}
      \caption{Light curves for MWC~560. From top to bottom: AAVSO+ASAS and
         $JHKL$ light curves, and $J-K$ colours, respectively.}
        \label{Figlc}
   \end{figure}
%

$JHKL$ broad-band photometry (1.25, 1.65, 2.2, 3.45 $\mu$m) was obtained
with the MkII infrared photometer on the 0.75 cm telescope at SAAO,
Sutherland (see \citeauthor{Carter1990} \citeyear{Carter1990} for details
about the system). The measurements are good to $\pm 0.03$ at $JHK$ and $\pm
0.05$ at $L$. The optical light curve consists of visual magnitude estimates
collected by the American Association of Variable Stars Observers (AAVSO),
and $V$-band photometry obtained in the frame of All Sky Automated Survey
(ASAS)
\citep{Pojmanski2002}. The near-IR observations cover the period from
November 1984 to February 2004 (92 points in $J$ and $K$ bands, 91 in $H$
and 86 in $L$) whereas the optical photometry was collected between April
1990 and December 2001 - the AAVSO data (2051 points), and from November
2000 to February 2006 - the ASAS data (330 points), respectively. The
near-IR data are listed in Table~1 and the light curves are presented in
Fig.~1.
\begin{table}
\caption{IR photometry of MWC~560}
\centering 
\begin{tabular}{c c c c c}   
\hline\hline
 JD & $J$ & $H$ & $K$ & $L$\\
 -2\,440\,000 & & & & \\
 (day) & \multicolumn{4}{c}{(mag)} \\
\hline
 6024.55 & 6.50 & 5.49 & 5.12 &       \\
 6026.45 & 6.51 & 5.50 & 5.14 & 4.74 \\
 6034.53 & 6.48 & 5.49 & 5.12 & 4.79 \\
 6043.57 & 6.49 & 5.48 & 5.13 & 4.79 \\
 6082.46 & 6.54 & 5.52 & 5.16 & 4.84 \\
 6108.39 & 6.54 & 5.53 & 5.18 & 4.84 \\
 6113.33 & 6.56 & 5.56 & 5.19 & 4.86 \\
 6136.30 & 6.54 & 5.52 & 5.16 & 4.85 \\
 6184.24 & 6.49 & 5.44 & 5.10 & 4.90 \\
 6338.57 & 6.50 &       &       &       \\
 6380.61 & 6.49 & 5.45 & 5.12 & 4.73 \\
 6440.46 & 6.59 & 5.54 & 5.20 & 4.90 \\
 6454.48 & 6.56 & 5.53 & 5.17 & 4.89 \\
 6465.35 & 6.54 & 5.50 & 5.16 &       \\
 7958.39 & 6.43 & 5.45 & 5.11 & 4.76 \\
 7961.37 & 6.42 & 5.44 & 5.11 & 4.80 \\
 7964.33 & 6.43 & 5.43 & 5.10 & 4.75 \\
 7989.26 & 6.44 & 5.42 & 5.09 & 4.78 \\
 8023.22 & 6.42 & 5.41 & 5.06 &       \\
 8028.19 & 6.39 & 5.42 & 5.08 & 4.71 \\
 8197.59 & 6.47 & 5.49 & 5.16 & 4.84 \\
 8227.56 & 6.48 & 5.49 & 5.16 & 4.79 \\
 8252.56 & 6.48 & 5.49 & 5.15 & 4.82 \\
 8280.49 & 6.43 & 5.44 & 5.11 & 4.71 \\
 8294.38 & 6.43 & 5.45 & 5.11 & 4.74 \\
 8320.40 & 6.44 & 5.44 & 5.10 & 4.80 \\
 8326.41 & 6.47 & 5.45 & 5.11 & 4.79 \\
 8390.20 & 6.44 & 5.43 & 5.08 & 4.74 \\
 8617.53 & 6.44 & 5.42 & 5.08 & 4.77 \\
 8670.41 & 6.48 & 5.41 & 5.07 & 4.69 \\
 8705.33 & 6.49 & 5.41 & 5.05 & 4.70 \\
 8731.29 & 6.42 & 5.37 & 5.03 & 4.69 \\
 8768.20 & 6.45 & 5.44 & 5.09 & 4.74 \\
 8899.63 & 6.49 & 5.48 & 5.14 & 4.82 \\
 8932.56 & 6.47 & 5.45 & 5.15 & 4.85 \\
 8960.52 & 6.44 & 5.43 & 5.10 & 4.73 \\
 8968.44 & 6.42 & 5.41 & 5.06 & 4.71 \\
 8969.52 & 6.42 & 5.43 & 5.09 & 4.79 \\
 8991.41 & 6.44 & 5.40 & 5.06 & 4.70 \\
 9024.42 & 6.43 & 5.41 & 5.06 & 4.77 \\
 9292.54 & 6.47 & 5.43 & 5.09 & 4.79 \\
 9379.36 & 6.38 & 5.36 & 5.01 & 4.67 \\
 9503.19 & 6.52 & 5.50 & 5.17 & 4.80 \\
 9643.62 & 6.53 & 5.52 & 5.15 & 4.83 \\
 9667.53 & 6.51 & 5.49 & 5.15 & 4.82 \\
 9728.50 & 6.47 & 5.44 & 5.10 & 4.72 \\
10035.54 & 6.48 & 5.47 & 5.13 & 4.81 \\
10053.60 & 6.46 & 5.44 & 5.10 & 4.73 \\
10086.47 & 6.50 & 5.48 & 5.13 & 4.79 \\
10110.42 & 6.52 & 5.49 & 5.14 & 4.77 \\
10126.45 & 6.52 & 5.50 & 5.15 & 4.87 \\
10181.29 & 6.46 & 5.46 & 5.11 & 4.74 \\
10221.22 & 6.45 & 5.43 & 5.09 & 4.72 \\
10437.54 & 6.45 & 5.44 & 5.10 & 4.78 \\
10464.46 & 6.49 & 5.45 & 5.11 & 4.79 \\
10478.37 & 6.49 & 5.46 & 5.12 & 4.81 \\
10498.39 & 6.48 & 5.47 & 5.14 & 4.86 \\
\hline                   
\end{tabular}
\end{table}
\begin{table}
\setcounter{table}{0}
\caption{continued.}
\centering
\begin{tabular}{c c c c c}        
\hline\hline
 JD & $J$ & $H$ & $K$ & $L$\\
 -2\,440\,000 & & & & \\
 (day) & \multicolumn{4}{c}{(mag)} \\
\hline
10567.23 & 6.50 & 5.47 & 5.14 & 4.77 \\
10801.51 & 6.54 & 5.49 & 5.14 & 4.81 \\
10884.38 & 6.47 & 5.42 & 5.09 &       \\
11146.59 & 6.48 & 5.47 & 5.11 & 4.86 \\
11187.48 & 6.51 & 5.48 & 5.13 & 4.72 \\
11234.30 & 6.46 & 5.43 & 5.08 &       \\
11479.54 &       &       & 5.14 &       \\
11505.51 & 6.53 & 5.48 & 5.13 & 4.77 \\
11511.48 & 6.52 & 5.46 & 5.12 & 4.81 \\
11572.48 & 6.52 & 5.46 & 5.11 &       \\
11599.35 & 6.50 & 5.46 & 5.12 & 4.82 \\
11611.33 & 6.52 & 5.48 & 5.12 & 4.78 \\
11617.34 & 6.52 & 5.49 & 5.15 & 4.76 \\
11675.24 & 6.50 & 5.46 & 5.11 & 4.76 \\
11857.58 & 6.54 & 5.52 & 5.17 & 4.86 \\
11864.58 & 6.55 & 5.54 & 5.19 & 4.90 \\
11869.52 & 6.56 & 5.52 & 5.18 & 4.87 \\
11881.54 & 6.54 & 5.51 & 5.18 & 4.87 \\
11890.57 & 6.55 & 5.52 & 5.18 & 4.85 \\
11927.47 & 6.54 & 5.52 & 5.18 & 4.88 \\
11976.31 &       &       &       & 4.85 \\
12239.58 & 6.49 & 5.47 & 5.13 & 4.75 \\
12255.57 & 6.53 & 5.49 & 5.15 & 4.80 \\
12263.47 & 6.51 & 5.50 & 5.16 & 4.75 \\
12284.50 & 6.52 & 5.49 & 5.13 & 4.76 \\
12289.44 & 6.51 & 5.48 & 5.16 & 4.81 \\
12319.40 & 6.48 & 5.45 & 5.12 & 4.74 \\
12324.44 & 6.49 & 5.47 & 5.10 & 4.80 \\
12349.32 & 6.48 & 5.44 & 5.10 & 4.80 \\
12384.26 & 6.49 & 5.43 & 5.09 & 4.85 \\
12572.58 & 6.59 & 5.53 & 5.17 & 4.87 \\
12594.60 & 6.56 & 5.52 & 5.15 & 4.81 \\
12664.43 & 6.52 & 5.50 & 5.15 & 4.87 \\
12690.38 & 6.52 & 5.45 & 5.09 & 4.79 \\
12979.55 & 6.49 & 5.45 & 5.12 & 4.79 \\
13031.48 & 6.48 & 5.41 & 5.06 & 4.73 \\
13063.34 & 6.41 & 5.37 & 5.03 & 4.66 \\
\hline                        
\end{tabular}
\end{table}
\section{Period Analysis}

We analysed the near-IR, ASAS, and combined AAVSO+ASAS magnitudes by means of the
program PERIOD\footnote{source of program is available on {\it
http://www.starlink.rl.ac.uk/}} ver. 5.0, which uses the modified
Lomb-Scargle method \citep{PressandRybicki1989}.  To remove long-term trends from the optical data,
a second order polynomial was subtracted from the combined AAVSO+ASAS light curve,
and a third order spline function was subtracted from the ASAS light curve.
The resulting power spectra are shown in the right panel of Fig.~2, and
results of our period analysis are summarized in Table 2. The periods given
in the table were derived from the maxima of the peaks in the
periodograms ($f_{max}^{-1}$), whereas their accuracy was estimated by
calculating the half-size of a single frequency bin ($\Delta f$), centred on
the peak ($f_{c}$) in a periodogram, and then converted to period units
($\Delta P = f_{c}^{-2} \cdot \Delta f$).

Power spectra of the composite  AAVSO+ASAS light curve (Fig.~2, the
uppermost, right panel) show a very pronounced peak at $1931^{\rm{d}}$
($P_\mathrm{act}$) very close to the period reported by
\citet{Doroshenko1993}. The second and third harmonics,
$\sim$1931/2 and $\sim$1931/3, respectively, are also present. 
We attribute this periodicity to repeated episodes of enhanced activity of the hot component following the ephemeris:
$${\rm{JD(max)}}=2\,448\,080 + 1931\times E$$
The light curve folded with this  period is also show in Fig.~2 (the
uppermost, left panel).

Power spectra corresponding to the $JHKL$ light curves are very similar to
each other (Fig. 2).  In all spectra the strongest peak corresponds to a
339-day period ($P_\mathrm{pul}$) which we take to be a real periodicity
produced by pulsations of the cool component. There are also peaks around
$\sim$4500-5500~days (which is comparable to the time interval spanned by
the data) connected with the long-term trend. Power spectra of the $HK$ light
curves show also peaks at $\sim1900^{\rm{d}}$ days and $\sim 310^{\rm{d}}$.
The former is very close to
$P_\mathrm{act}$ found in the optical light curve whereas the latter seems
to be an annual alias of 1900-day peak rather than a real periodicity. In
the power spectrum of the $J$ light curve the $\sim$1900-day and
$\sim$310-day periods are not seen because they are overwhelmed by the strong
long term trend. These periods are absent in the power spectrum of the $L$
light curve. The following ephemeris gives the phases of maxima in the
$JHKL$ bands:
$${\rm{JD(max)}}=2\,445\,960 + 339\times E$$
and the light curves folded with this ephemeris are shown in Fig.~2. (left panel).

The power spectra of the visual photometry also show peaks at $\sim
310^{\rm{d}}$.  These power spectra have poorer resolution that those for
the near-IR data, so the peaks, especially that for the ASAS data, are
relatively broad, and they may in fact be due to the $339^\mathrm{d}$
pulsation period combined with one year alias of the
$1931^\mathrm{d}$ period. Such an interpretation is supported by the
presence in the ASAS power spectrum of another peak at $\sim
166^\mathrm{d}$. This peak can be more accurately determined and it is due
to the second harmonic of the pulsation period, which in this case would be
$332^\mathrm{d}$.  The ASAS light curve folded with this period is also
shown in Fig.~2.
 
In  the power spectrum  of the AAVSO+ASAS data  there is also a strong peak
at $747^{\rm{d}}$ connected with a quasi-periodic oscillation which is
predominantly visible during JD 2\,450\,500-2\,452\,300 (1997-2002). 
However, we note that it is very close to 2 years, and it may be an
artifact.

%
\begin{table}
\caption{Peaks in power spectra}             
\label{table:2}      
\centering                          
\begin{tabular}{c c c c}        
\hline\hline                 
Frequency & Period & Power & Remarks \\    
days$^{-1}$ & days & sigma unit & \\ 
\hline                        
J&&&\\
2.9479 10$^{-3}$  &  339$\pm$4 & 20.8 & $P_\mathrm{pul}$ \\
&&&\\
H&&&\\
5.3276 10$^{-4}$  & 1877$\pm$126 &  7.0 & $P_\mathrm{act}$\\
2.9479 10$^{-3}$  &   339$\pm$4  & 19.8 & $P_\mathrm{pul}$ \\
3.2321 10$^{-3}$  &   309$\pm$3  & 10.6 & alias\\
&&&\\
K&&&\\
5.3276 10$^{-4}$  & 1877$\pm$126 &  7.2 & $P_\mathrm{act}$\\
2.9479 10$^{-3}$  &  339$\pm$4   & 20.2 & $P_\mathrm{pul}$ \\
3.2676 10$^{-4}$  &  306$\pm$3   & 12.2 & alias\\
&&&\\
L&&&\\
2.9487 10$^{-3}$  &   339$\pm$4  & 11.6 & $P_\mathrm{pul}$ \\
&&&\\
ASAS&&&\\
3.2708 10$^{-3}$  &   306$\pm$12 & 35.8 & $P_\mathrm{ASAS}$ \\
6.0183 10$^{-3}$  &   166$\pm$4  & 18.5 & $P_\mathrm{pul}/2$ \\
&&&\\
AAVSO+ASAS&&&\\
5.1789 10$^{-4}$  & 1931$\pm$162 & 285.6 &  $P_\mathrm{act}$\\
9.4946 10$^{-4}$  & 1053$\pm$48  &  57.9 &  $P_\mathrm{act}$/2\\
1.3379 10$^{-3}$  & 747$\pm$24   & 102.4 &  \\
3.1937 10$^{-3}$  & 313$\pm$4    &  53.1 &  $P_\mathrm{ASAS}$, alias\\
\hline                                   
\end{tabular}
\end{table}


   \begin{figure}
   \centering
   \includegraphics[width=8.5cm]{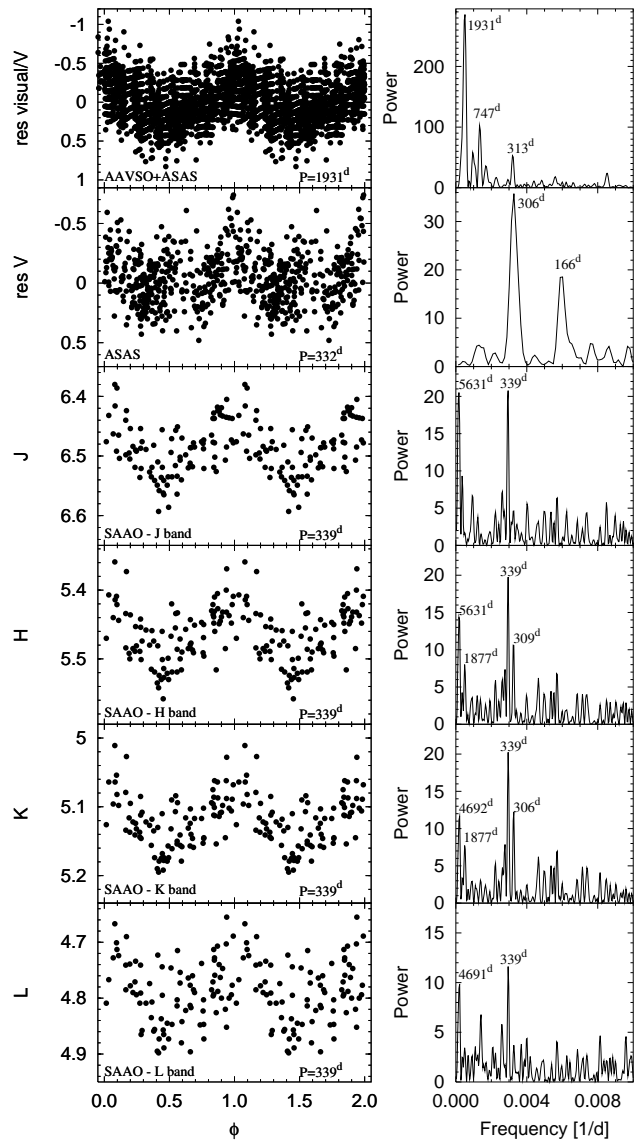}
   \caption{Power spectra (right) and corresponding light curves folded with the strongest period (left). From the top to bottom: AAVSO+ASAS light curve folded with period $P_\mathrm{act}$=1931 days,   ASAS light curve folded with period $P_\mathrm{ASAS}$=332 days, and $JHKL$ light curves all folded with period $P_\mathrm{pul}$=339 days, respectively.
 }
              \label{Figfaza}%
    \end{figure}

\section{Discussion and conclusions}

Our near-IR photometry began in 1984, with only a dozen measurements over a
$~1.5$-year interval.  The observations restarted in 1990, when MWC 560
underwent an outburst, reaching the brightest level in its whole photometric
history, and ejected jets \citep{Toma1990}. In the near-IR, it brightened by
$\sim$0.1~mag in all, $JHKL$, bands. Then during the following 12 years, it
slowly faded to the pre-outburst magnitudes observed in 1984. In 2002-2004,
the star brightened again.  The point scatter of the near-IR light curves
was always of $\sim$0.1~mag in $JHK$ bands, and $\sim$0.2~mag in $L$ band,
respectively. In general, the near-IR light curves reflect the trend shown
by the visual/$V$ light curve, with additional maxima around JD 2\,450\,000
and JD 2\,452\,000. The $J-K$ colour becomes redder as the brightness
declines (Fig. 1) due to a decreasing contribution from the hot component.

Our period analysis of the $JHKL$ light curves revealed a 339-day
periodicity which can be attributed to radial pulsations of the M giant.
Although this period is close to that of Mira itself, and of many other
Galactic Mira variables, the amplitude of the pulsation, $\Delta\,K
\sim 0.1$~mag, is much lower than $\Delta\,K \ga 0.4$~mag observed in
Miras, including typical symbiotic Miras \citep{Whitelock1987}. We therefore
classify it as an SRa variable (using the definition in the GCVS4) rather
than as a Mira.  This pulsation is hardly detectable in the visual light
because radiation in this range is dominated by the very active hot
component. In fact, the scatter of points from the folded light curve in
Fig. 2, $\Delta V
\sim 0.3 \div 0.6$ mag is comparable to the amplitude of the flickering
\citep{grom2006} which complicates the period analysis. The red giant
pulsation period found in this study is more than twice the 161-day
pulsation period reported by \citet{Franckowiak2003}, and it is the only
periodicity present in all bands.  

The basic properties and evolutionary status of the semi-regular variables
(SRVs) of type SRa and SRb were discussed in detail by
\citet{Kerschbaum_and_Hron1992, Kerschbaum_and_Hron1994,
Kerschbaum_and_Hron1996}. In particular, they found that the SRas appear as
intermediate objects between Miras and SRbs in all aspects, including
periods, amplitudes and mass-loss rates. They also concluded that the SRas
do not form a distinct class of variables, but are a mixture of `intrinsic'
Miras and SRbs. The SRbs split into a `blue' group with $P<150$ days and no
indication of circumstellar shells and a `red' group with temperatures and
mass-loss rates comparable to Miras, but periods about half as long. They 
suggested that the `red' and `Mira' SRbs are thermally pulsing
AGB-stars \citep{Kerschbaum_and_Hron1992}.  The persistent and relatively
long period places the cool component of MWC 560 among the `Mira' SRVs.
These differ from normal Miras only in their smaller pulsation amplitudes. 
The IRAS [12\,$\mu$m]-[25\,$\mu$m]=0.64
locates MWC 560 in the period--IRAS colour diagram of
Kerschbaum \& Hron (1992) in the region occupied by 86 \% of the Miras,
whereas most SRbs are outside this region. The average, reddening corrected
($E_\mathrm{B-V}=0.15$; Schmid et al. 2001) near-IR colours $\langle H-K
\rangle_0=0.30$, $\langle J-H \rangle_0=1.00$, $K-[12\,\mu$m]=1.25, and the
spectral type M5.5$-$6 are also consistent with an O-rich SRV, although the
colours are unlike those of a Mira. Finally, if the 1931 day periodicity is
orbital then the two stars of this symbiotic system are close to each other
(see below), and the presence of a nearby companion may influence the
pulsation characteristics, possibly reducing the pulsation amplitude. We
suggest that the evidence supports the view that the red component of
MWC~560 is on the TP AGB-phase.

 The SRV component of MWC 560 could be the source of a strong stellar wind
and thus support the observed high activity of the hot component. If we
estimate the mass of the red giant at $1M_{\odot}$ and that of the white
dwarf companion at $0.5M_{\odot}$ then, assuming that the 5.3 year period is
orbital, the separation of the two stars will be 3.55 AU and the radius of
the Roche lobe for the M giant in a circular orbit will be 1.5 AU.
\citet{Schmid2001} argued that an accretion rate up to $\rm {a\,few} \times
10^{-7}\, M_{\sun}\, \mathrm{yr}^{-1}$ is necessary to account for the
observed hot component luminosity which is comparable to the wind efficiency
found for typical Miras and SRVs. For example, \citet{Olofsson2002}
determined average mass loss rate of $2 \times 10^{-7}\, M_{\sun}\,
\mathrm{yr}^{-1}$, with a maximum value of $8 \times 10^{-7}\, M_{\sun}\,
\mathrm{yr}^{-1}$ for a sample of M-type irregular and semi-regular
variables. So, in the case of MWC 560, a significant fraction of the mass
lost in the cool giant wind must be accreted by the hot companion. This is
possible only if the binary components of MWC 560 interact via `wind Roche
lobe' overflow occurs instead of a spherically symmetric wind
\citep{Podsiad2006}.

One of the most intriguing features of this hot component activity is its
highly periodic character. The 1931-day periodicity found in the visual
AAVSO+ASAS light curve is essentially identical with the 1930-day
periodicity detected in the $m_\mathrm{pg}/B$ light curve by
\citet{Doroshenko1993}. We note that this periodicity remained in-phase over
a century, and the most natural explanation would be an orbital motion. The
only problem with the orbital interpretation is that the orbital inclination
of MWC~560 seems to be extremely low (as indicated by the jet axis
practically aligned with the line of sight, and the lack of orbitally
related radial velocity changes) which excludes any geometrical effects, such
as eclipses, illumination, etc.  Although one can argue that the jet axis
can be inclined with respect to the binary orbital plane, the relatively
large amplitude of this variability, $\Delta m_\mathrm{pg} \sim 1$ and
$\Delta V \sim 0.5$ mag, respectively, points to another mechanism(s).
One possibility is an eccentric orbit and enhanced accretion rate
near periastron passage causing a brightening of the hot component
\citep{Doroshenko1993}. Then we can even speculate that if the eccentricity is high enough, $\sim 0.5$ or more,  a Roche lobe overflow may occur near the periastron.

Another possibility is enhanced mass loss due to periodic changes in the SRV
environment.  We note that the 1931~day period equals roughly six times the
pulsation period.  A periodically enhanced mass loss might be caused by some
kind of interplay between the cool component pulsation and layered dust
formation shells. Such effects have been shown by time-dependent
hydrodynamic simulation of C-rich Mira environments
\citep{Winters1999, Winters2001}.  In particular, these simulations have
been able to reproduce variations of the mass-loss rates on different time
scales.

\begin{acknowledgements}
This study made use of the 
      American Association of Variable Star Observers (AAVSO)
      International Database contributed by observers worldwide and the public domain database of The All Sky Automated Survey (ASAS) which we acknowledged. 
We also thank the following people for making IR observations: Greg Roberts, Robin Catchpole, Brian Carter, Dave Laney and Hartmut 
Winkler.
This research was partly supported by KBN grant 1P03D\,017\,27.
\end{acknowledgements}

\bibliography{6538ref}        
\bibliographystyle{aa}

\end{document}